\documentclass[journal]{IEEEtran}

\usepackage[final]{changes}

\usepackage{amsthm}
\usepackage{graphicx}

\usepackage{url}
\usepackage{balance}
\usepackage{cite}

\usepackage{subfigure}

\usepackage{comment}

\usepackage{color, soul}

\usepackage{cancel}

\usepackage{amssymb}
\newtheorem{thm}{Theorem}

\begin{document}

%

\title{Analog Neural Computing with Super-resolution Memristor Crossbars}

%
%

\author{Alex James,~\IEEEmembership{Senior Member,~IEEE,}
        and~Leon~Chua,~\IEEEmembership{Fellow,~IEEE}
}

\maketitle


\begin{abstract}
Memristor crossbar arrays are used in a wide range of in-memory
and neuromorphic computing applications. However, memristor devices suffer
from non-idealities that result in the variability of conductive
states, making programming them to a desired analog conductance value extremely difficult as the device ages. In theory, memristors can be a nonlinear programmable analog resistor with memory properties that can take infinite resistive states. In
practice, such memristors are hard to make, and in a crossbar, it is confined to a limited set of stable conductance values. The number of conductance levels available for a node in the crossbar is defined as the crossbar's resolution. This paper presents a technique to improve the resolution \deleted{of the crossbar} by building a super-resolution memristor crossbar with nodes having multiple memristors to generate $r$-simplicial sequence of unique conductance values. The wider the range and number of conductance values, the higher the crossbar's resolution. This is particularly useful in building analog neural network (ANN) layers, which are proven to be one of the go-to approaches for forming a neural network layer in implementing neuromorphic computations.
\end{abstract}%

\begin{IEEEkeywords}
Crossbar, analog neural network, multiply and accumulate, analog neural accelerator
\end{IEEEkeywords}

%
\IEEEpeerreviewmaketitle


\section{Introduction}

{\label{640088}}

\IEEEPARstart{I}{n} the design of neural network accelerators, multiply and accumulate (MAC)\cite{Zhang_2020,Camus_2019,Kang_2020} block emulates the weighted summation of inputs in a neural network layer. Every layer of the neural network requires the MAC operation. The neural network accelerators \cite{Tann_2018,Siu_2018,Seto_2019,Yoo_2019,Wen_2019} are
used in the edge AI computing domain to make sensing smarter and data processing energy efficient. The offloading of neural network computing from high-performance computers to edge AI processors improves energy efficiency, and reduces costs, and bandwidth requirements.

The MAC block can be implemented with digital or analog circuits. In a digital-only approach, the inputs, and weights are stored in a binary form. Such a system is most suitable if using binary memory arrays, such as SRAM or DRAM \deleted{,b for implementing MAC} through in-memory computing \cite{Yin_2020,Jiang_2020,Choi_2020}. In an analog approach, the MAC  operation uses  weights and/or inputs in analog forms, such as when implementing
using a floating gate or ReRAM array\cite{Liu_2020}. The most popular approach to building MAC in the analog domain is \deleted{y using a} \added{the} crossbar architecture\cite{Starzyk_2014,Yakopcic_2017}.

In a crossbar architecture, the crossbar nodes can be accessed by row or column lines. The rows are used to apply the input signals, while the columns are used to read the output signals. {The crossbar nodes of our interest have a programmable conductance. When a voltage is applied to the crossbar node, the column's output current represents the weighted summation of inputs, or rather the MAC operation.}

A popular conductance device for building a crossbar node is the \deleted{memristors} \added{memristor}\cite{Li_2018,Sheridan_2014}. \deleted{There are a large variety of}\added{Several} devices \deleted{that} claim to have similar properties as memristors\cite{prodromakis2010review,sheridan2019memristors, milano2019nanowire, zhu2020comprehensive}. {The neural network applications of memristors in the crossbar require the mapping of neural network weights to the crossbar nodes' conductance values.} This mapping of weight is only possible on a single memristor \deleted{if the memristor has}\added{having} infinite \deleted{number of} conductance levels. \deleted{As}\added{Because} memristors with precise and stable infinite conductance
levels are currently not practically feasible, encoding schemes \deleted{needs}\added{need} to be applied.

One approach is to encode the weights as a binary sequence or to discretize the weights to fit the available \deleted{number of} \added{memristor} conductance levels \cite{Ni_2017,Sebastian_2020}. This limitation of weight resolutions \deleted{of the memristor} determines the overall design of the MAC implementation, and thereby the neural network performance. Subsequent to the resolution limitation, the memristor is also sensitive to process variations and device-to-device variabilities, which leads to
conductance variability \cite{gale2015non, jeong2017parasitic, krestinskaya2019memristive,vourkas2015alternative,chua2019handbook}. The endurance and aging of the device play an
important role in the reliability of the crossbar, and often make the circuit designers work difficult~\cite{Zhang_2019}. 

{Connecting memristors in parallel using PCM devices  \cite{bill2014compound,boybat2018neuromorphic}  increases the reliability and fault tolerance of synapses. Likewise, ReRAM based parallel memristors were used in \cite{boybat2019multi} to reduce the impact of stochasticity and nonlinearity. Whereas, in this paper, we present the theory of combining the memristors within a crossbar without modifying the widely used crossbar structures as a fundamental technique to generalise crossbar for any precision multiply and accumulate operation.  A practical approach to improve the conductance resolution of crossbar arrays is proposed.} The main contributions of this work include (1) a super-resolution crossbar design, (2) theory of estimating the conductance levels and (3) implication on the practical design of {analog} neural
networks.

This paper is organized into the following sections: Section II gives a background on \deleted{human brain and} crossbars that inspire analog computing, Section {\ref{437585}} provides the details of the proposed design approach and \deleted{the} supporting theory, Section {\ref{460767}} reports the results and \added{related} discussion\deleted{of the results}. Finally, Section {\ref{301895}} provides the summary of the paper.

\section{Background}

\begin{figure}
    \centering
    \includegraphics[width=60mm]{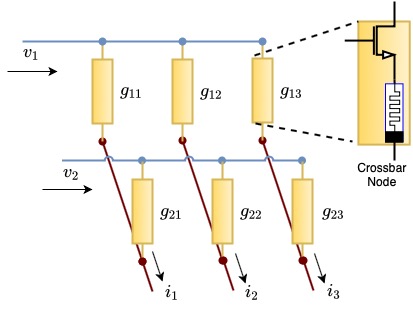}
    \caption{{The memristor crossbar with $v_1,v_2$ as inputs, conductance  \added{$g_{i,j}$} of memristors as weights and columns current $i_1,i_2,i_3$ as outputs. Here, the node shown is formed with a transistor switch and memristor (1T1M).  }}
    \label{fig:figc}
\end{figure}


{\label{549990}}\par\null

{The crossbar arrangement of memristors in 1T1M configuration is shown in Fig.
\ref{fig:figc}. The transistor switch enables or disables the column-wise memristor nodes.} In this example, the inputs ($v_1,v_2$) are applied across the rows of the crossbar, and outputs ($i_1,i_2,i_3$) measured at the columns of the crossbar. Each memristor has a conductance $g_{ij}$, where $i,j$ are the coordinates of the crossbar nodes. The output currents indicate the weighted summation of inputs. {The output currents from the $n$th neuron (column of crossbar) can be represented as $i_n=\sum_{k=1}^N v_kg_{k,n}$, with $N$ number of inputs (number of rows in the crossbar), representing a dot-product or MAC operation or vector-multiplication (VMM) \cite{Miranda2020,Hu2018,Kim2017,Sheridan2017,Ngoc2017,Azghadi2017,Kim2016,Zidan2016,Truong2015}.} In  a  neural  network implementation, this weighted summation of inputs represents a neuron representation without activation function.  {In the crossbar, bias can be included by adding additional input line. The memristor programming and line selection is done using a selector switch in series with the memristor as shown in Fig. 1.} {The selector switch enables the memristor node by applying a gate voltage sufficiently higher than the threshold voltage to drive the transistor switch to ON mode. } {In analog neural networks, the  $g_{i,j}$ are weight variables that can take continuous values. In most practical hardware implementations, the $g_{i,j}$ is a discrete variable \added{having }limited set of values. As memristors have high readout speeds in crossbar, the MAC operations can \deleted{be used for performing for}\added{perform} analog numerical computations \cite{chen2021multiply1}. An example is the \deleted{system that used} Phase Change Memory based memristor crossbar to solve linear system of equations using iterative refinement algorithm \deleted{and} in a mixed-signal architecture. Such a system showed {computational} complexity of $O(\textnormal{D}^2)$ for an D $\times$ D matrix, \deleted{in comparison}\added{compared} to $O(\textnormal{D}^3)$ in traditional
numerical computation algorithms \cite{le2018mixed}.}

\deleted{Considering the inputs $x$ and weights $w$ as a vector, the output
is simply the dot-product between these two variables. This represents the layer of the neural network's dot-product computation, which is emulated by the crossbar\cite{Miranda2020,Hu2018,Kim2017,Sheridan2017,Ngoc2017,Azghadi2017,Kim2016,Zidan2016,Truong2015}.
In addition to the weight, one more additional input line is added for bias in the crossbar. The memristor programming and line selection can be made using a selector switch in series with the memristor.}

\deleted{Similar to memristors, transistor-based crossbars are also used to build neural computations. An example of this is the IBM TrueNorth chip \cite{Fair2019} used for spiking neural network implementations. However, the bottleneck is the high cost and poor area efficiency.}

The crossbars' reading and writing require additional circuits specific to the neural network architecture. In many neural networks, the weights \deleted{are composed of}\added{can be} positive and negative\deleted{ numbers}. To implement such a system, each synaptic column is represented as two distinct column lines, thereby doubling up the number of memristors required. {The negative and positive signs are achieved by splitting the conductance components as positive $g_{i,+}^+$ and negative $g_{i,-}^-$ along two crossbar columns (say positive and negative column), which reflects as $i_1^+-i_1^-=\sum_{i=1}^N v_i(g_{i,+}^+-g_{i,-}^-)$. Here, the weights of the neuron $w_i \propto (g_{i,+}^+-g_{i,-}^-) $, where $g_{i,+}^+$, and $g_{i,-}^-$ are conductance of memristor nodes in positive and negative columns, respectively, while current difference $i_1^+-i_1^-$ is realised with a sense amplifier having a current difference circuit.}

The readout circuits \deleted{usually will have}comprises sense amplifiers and data converters~\cite{Mohan_2019}. There are two main possibilities in processing the current outputs. {They can be passed through an analog implementation of the activation function followed by the next neural crossbar layer.} Alternatively, the output currents can be \deleted{converted to digital encoding}\added{digitally encoded} using a data converter, followed by activation function computation and the next layer of input. In an analog-only approach, the design of sense amplifier and activation function becomes challenging, as they are power-hungry, although it can be area efficient. In digital-only implementations, the power is on the lower end, but they become area inefficient due to data converters and multiplexing units.

\begin{figure}
    \centering
    \includegraphics[width=60mm]{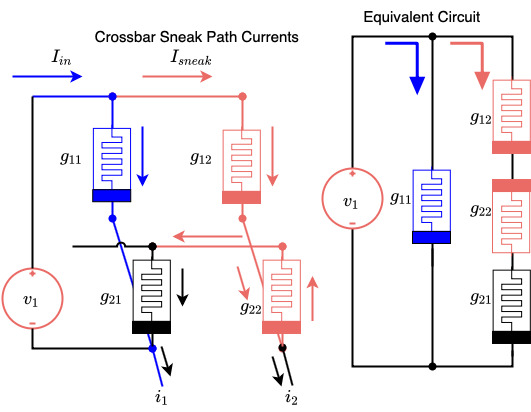}
    \caption{{Unintentional flow of current in $g_{12}$, $g_{22}$ and $g_{21}$ is representative of sneak path currents in a crossbar.}}
    \label{fig:sneekpath}
\end{figure}

\deleted{{Figure \ref{fig:sneekpath} shows \deleted{the}\added{an} example of sneak path currents in a crossbar.} The memristor nodes without row selector switches will have unintentional flow of currents sneaking from \added{the }neighbourhood columns to the readout column (see \cite{shi2020research}). In the example \deleted{illustrated} in Fig. \ref{fig:sneekpath} the unintentional flow of current occur in $g_{12}$, $g_{22}$ and $g_{21}$, which introduces the readout current errors in $i_1$.} {In Fig. 2 example, it can be noted that each column is assumed to be coupled to a sense amplifier. If the column with $i_2$ is not coupled to a sense amplifier, all current in such case will go to $i_1$. } The sneak path currents \cite{Cassuto_2013,Shi_2020,G_l_2019,Cassuto_2016,Naous_2014,Kannan_2013,Shevgoor_2015}
in the crossbar can be controlled by serial reads, \deleted{which makes}\added{making} the crossbar reads slower. {In serial read, one column in the crossbar is enabled at a time and the output currents are read along the column. In parallel read, multiple columns in the crossbar are enabled and the output currents are read along the enabled columns in parallel simultaneously.} A parallel read \added{is faster, but} \deleted{will} have more current errors\deleted{, although the reads will be much faster}. Smaller crossbars will have a fewer \deleted{number of} paths in the crossbar to induce sneak path errors. Keeping \added{the }crossbar size smaller also \deleted{has the advantages of lowering}\added{lowers} the line capacitance and signal integrity issues.

Increasing the crossbar size is required for building large neural networks. One way to get around this problem is \deleted{by} using modular crossbar arrays or crossbar tiles \cite{Mountain_2018,Mikhailenko_2018}. The current summation can be split into several sub-currents by splitting the large crossbar into a set of smaller crossbars followed by summing the output currents. As smaller crossbar will have smaller leakage currents, \added{lower sneak-path currents due to }the use of multiple crossbars to represent larger \added{crossbars}\deleted{crossbar results in lower sneak-path currents}, and parasitic errors.

\section{Super-resolution Crossbar
Nodes}

{\label{437585}}

\deleted{The crossbar consists of a set of inputs~\(v_i\) connected to the \(i\) rows and outputs \(i_k\) read as currents of the crossbar \(k\) columns. Each node in the crossbar has a memristor node having a conductance of \(g_{ik}\). The output current~\(i_k=\sum_{i=1}^Ng_{ik}v_i\), where \(N\) is the total number of inputs. In a neural network implementation, this weighted summation of inputs represents a neuron representation without activation function.} {In a memristor crossbar, any undesired change node conductance will lead to errors in output column current. Ideal memristors can be programmed to an infinite number of stable conductance values. However, practical memristors can be programmed to only a limited number of stable conductance values. The number of stable conductance values is defined as conductance levels.} If the conductance of the memristor is limited to \(L\) levels, i.e., \(L\) number of conductance values per device, many optimal weights of the neural network will become difficult to translate directly to hardware. One approach around this is to encode the weights with a set of memristor devices in the crossbar.

\begin{figure}[!ht]
\begin{center}
\includegraphics[width=60mm]{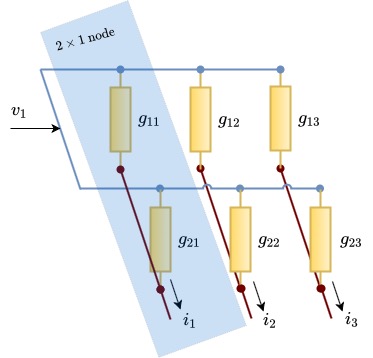}
\caption{{Memristor crossbar showing the proposed super-resolution conductance
nodes. In the illustration, each crossbar node is made up of two
memristors.
{\label{427153}}%
}}
\end{center}
\end{figure}

In {the} proposed approach, multiple rows are tied to a single input, as shown in
Fig. {\ref{427153}}, \deleted{which will enhance}\added{enhancing} the number of
conductance levels per node and increase the overall resolution of the crossbar. Suppose \deleted{we have} each memristor \deleted{having}\added{have} only two states, i.e. \(\left[g_l,\ g_h\right]\), then the weights are considered to be binary. Now, when we connect two rows for a single input, the overall conductance will be the summation of the two conductance, which means it
can take at most any of the three unique conductance values (see Table {\ref{table1}}). The occurrence of conductance combinations can be expressed in a polynomial \deleted{expression} form. If we represent, \(p_l\)  \deleted{as the symbolic occurrence of \(g_l\) ,} and \(p_h\) as the symbolic occurrence of  \added{\(g_l\) and} \(g_h\), \added{respectively, }a node with two memristors can be expressed as \(\left(p_{l+}p_h\right)^2=p_l^2+p_h^2+2p_lp_h\), where~\(p_l^2\) translates to \(2g_l\), \(p_h^2\) translates
to \(2g_h\), and \(2p_lp_h\) translates to \deleted{2}\(\left(g_l+g_h\right)\). Here, \(p_l\) and \(p_h\) are indicative of the number of conductance levels per memristor, and the power of 2 represents the number of memristors.

\begin{table}
\centering
\caption{Unique conductance values obtained by connecting two binary memristors in parallel to form a crossbar node.  }
\label{table1}
\begin{tabular}{ccc}\\\hline
Memristor 1& Memristor 2 & Overall {conductance}\\\hline
$g_l$& $g_l$ &$2g_l$\\
$g_l$& $g_h$ &$g_l+g_h$\\
$g_h$& $g_l$ &$g_l+g_h$\\
$g_h$& $g_h$ &$2g_h$\\\hline
\end{tabular}
\end{table}

Extending the idea to \(m\) memristors, this can be represented as \(\left(p_{l+}p_h\right)^m=\sum_{k=0}^m\frac{m!}{k!\left(m-k\right)!}p_l^kp_h^{m-k}\), \deleted{and that leads}\added{leading} to having
binomial properties for finding the unique coefficients. This indicates that the number of coefficients follows the simplicial polytopic numbers sequences.

\begin{thm}
{A combination of \textbf{\(m\)} bi-level conductance memristors in parallel \deleted{when used} for creating the super-resolution crossbar conductance node follows unique conductances (\(L_C\)) of {2,3,4..\textbf{\(m\)}+1} by encoding  1-simplicial number patterns. } {The increase in $L_C$ enables the conversion of binary conductance nodes to analog discrete conductance nodes in a crossbar.}
\end{thm}

{\textit{Proof:} Table \ref{table2} shows the number of memristors and corresponding unique conductance per node. Let's denote the unique conductance $(L_C)$ set for bi-level memristor nodes with $L=2$ as $U_g^2$. This sequence $U_g^2=\{2,3,..m+1\}$ are part of simplicial sequence if it encodes the 1-simplicial number properties. Here, we can rewrite, \\
$U_g^2=\{2,3,..m+1\}$ \\
$U_g^2=\{1,2,..m\}+\{1,1,..1\}$\\
$U_g^2=A+B$, for, $A=\{1,2,..m\}$ and $B=\{1,1,..1\}$.
Suppose, $t_{s(m)}$ represents the sum of $m$ numbers in the sequence $A$ and $t_m$ represents the sum of $m$ numbers in the sequence $U_g^2$. Here, $t_{s(m)}=\frac{m(m+1)}{2}$, and therefore, $t_m=t_{s(m)}-1$ for $m\ge 0$. For example,  for $m=2$, $t_{s(2)}=\frac{2(2+1)}{2}=3$; for $m=3$, $t_{s(3)}=\frac{3(3+1)}{2}=6$, and so forth, with $t_m=2$ for $m=2$, $t_m=5$ for $m=3$, and so forth. The conductance with different memristor numbers are given in Table \ref{table2}.
The sum of the sequence $t_{s(m)}$ validates the existence of 1-simplicial sequence or the Triangular gnomon (linear) number sequence.
We can also further observe additional relations between $t_m$ and $t_{s(m)}$, for computing the value of $m$ given $t_{s(m)}$ or $t_m$, and vice versa:\\
(1) $t_m-t_{m-1}=t_{s(m)}-t_{s(m-1)}=L_C(m), m\ge 1$,\\
(2) $t_m-2t_{m-1}+t_{m-2}=t_{s(m)}-2t_{s(m-1)}+t_{s(m-2)}=1, m\ge 2$,\\
 (3) $\frac{t_m}{t_{m-1}}=\frac{t_{s(m)}-1}{t_{s(m-1)}-1}$. \\
 Given that $t_{s(m)}$ validates $A$ for being 1-simplicial numbers and as it is encoded in $U_g^2$; we can say that $L_C$ for $L=2$ encodes 1-simplicial numbers. This proves Theorem 1 on the existence of 1-simplicial number patterns in the unique conductance set by combining bi-level memristors. $\square$
}

Theorem 1 has implications in the design and use of binary crossbar array. Any binary crossbar array can be transformed into an analog crossbar array using the proposed super-resolution principle.  In other words, the memristors with two states can \deleted{be used to} create nodes with \deleted{a larger number of}\added{several} conductance levels by adding more memristors in parallel. As
shown in Table~{\ref{table2}}, \deleted{by having} a node
with~\(m\) memristors\deleted{, it is possible to}\added{ can}
achieve~\(m+1\) conductance values. Because we use 10 rows of the crossbar to represent single input, i.e., ten memristors per input node, with each having a binary conductance state, we can obtain 11 unique conductive states, which \deleted{on}\added{in} itself does not show as a major advantage. However, with \added{the increasing}\deleted{increase in} number of conductive states per memristor\added{,} the scenario significantly changes. {For example, two memristors in a node with each memristor having two conductive states results in unique conductance ($L_C$) that follow 2-dimensional simplicial number sequence as shown in Table III.}

\begin{table}
\centering
\caption{Number of unique conductance values obtained from $m$ binary memristor per crossbar node. }
\label{table2}
\begin{tabular}{cc}\\\hline
Number of memristors& Unique {conductance}\\\hline
1&2\\
2&3\\
3 &4\\
4 &5\\
..&..\\
$m$&  $m+1$\\\hline
\end{tabular}
\end{table}

The~simplicial polytopic numbers belong to the broad class of figurate numbers that are \(r\)-dimensional simplex for each dimension \(r\) in the Euclidean space \(R^r \),
with \(r\ge0\).  In algebraic topology, the \(r\)- simplex is a \(r\)-dimensional polytope and is a convex hull of ~\(r+1\) vertices. They can be visualized as a generalization of the triangle to any arbitrary
dimension, \deleted{which can be} connected to the coefficients of multi-nominal expressions. These topological representations lay the foundations of the simplicial polytopic numbers.

\begin{figure}[!ht]
    \centering
    \includegraphics[width=60mm]{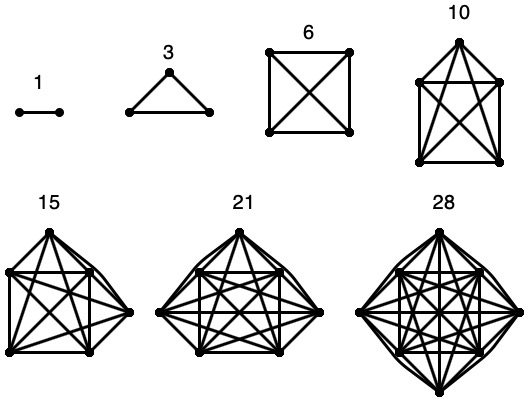}
    \caption{{Visualisation of 2-simplicial numbers as number of segments of triangles. With each addition of a vertex and counting the number of all connected segments gives simplicial polytopic number. }}
    \label{fig:tri}
\end{figure}

The number of unique conductance combinations follows multi-nominal expressions and can be represented as \(r\) simplicial numbers. 2-simplicial numbers originate out of 2-simplex having 3 vertices representative of the triangle (the 3 1-cell, with 3 segments as facets). {Figure \ref{fig:tri} shows the visualisation of the 2-simplicial numbers derived by counting the number of segments (or edges) of triangular shapes with an increasing number of vertices. Counting the number of segments in each shape gives the sequence,  $\{1,3,6,10,15,21,28...\}$, which forms the 2-simplicial numbers.} Likewise, 3-simplicial numbers originate out of 3-simplex having 4-vertices representative of the tetrahedron (the 4 2-cell, with 4 faces as facets). \deleted{And in}\added{In} general, \(r\)-simplicial numbers originate out of \(r\)-simplex having \(r+1\) vertices representative of the triangle (the \(r+1\) \(\left(r-1\right)\)-cell, with \(r+1\) segments as facets).  

\begin{thm}
The \textbf{ \(m\) \(L\)}-level conductance memristors super-resolution crossbar conductance node follows
an \(r\)-dimensional simplicial number sequence. {Increasing\textbf{~\(m\)}~irrespective of \textbf{\(L\)}~results in an increase in \(L_C\), such that \(L_C\ge L\).}
\end{thm}

{
\textit{Proof:} Table \ref{table3} shows the set of unique conductance ($L_C$)  following simplicial numbers for nodes with $m$ number of memristors. The conductance values for $L=\{1,2,3\}$, for $m=\{1,2,3\}$ is  shown in Table  IV, which follows the $\{1,2,3\}$-simplicial number sequences. For a simplicial sequence, the $k$th simplicial $r$-polytopic number can be calculated as: $S_{r+1}^{(r)}(k)={r+ (k-1) \choose k} =\frac{k^{(r)}}{r!}$. Replacing, $k$ with $m$, and $r$ with $L$, we find that the values of $L_C$ in Table  IV is same as that obtained by applying equation for $S$, as $\{1,2,3\}$, $\{1,3,6\}$ and $\{1,4,10\}$. Here, we observe that $L_C=f(m,L)$ is a function of $m$ and $L$, with:\\
$f(2,1)=f(2,0)+f(1,1)$\\
$f(2,2)=f(2,1)+f(1,2)$\\
$f(2,3)=f(2,2)+f(1,3)$\\
.......\\
$f(m,L)=f(m,L-1)+f(m-1,L)$\\
This implies that,  $L_{C_{L+1}}^{(L)}(m)$=$L_{C_{L+1}}^{(L)}(m-1)$+$L_{C_{L+1}}^{(L-1)}(m)$, is the recurrence relation for $m$th simplicial $L$-polytopic numbers. Applying the equation for $S_{r+1}^{(r)}(k)$ and replacing $k$ with $m$, and $r$ with $L$, the unique number of conductance sequence from a node with $m$ memristors can be then written as $\{ 1, \frac{m^{(1)}}{1!}, \frac{m^{(2)}}{2!}, \frac{m^{(3)}}{3!},..., \frac{m^{(L)}}{L!}\}$. Here, $L\ge 1$ can be  considered as the dimension of the regular convex simplicial polytope number sequence, while $m\ge 1$ is the number of memristors per node  as the number of nondegenerate layered simplices. For any given value of $m$, suppose the memristor has $L$ levels, then the unique number of levels $L_C$ becomes $\frac{m^{(L)}}{L!}$, which makes $L_C = L$ for $L=1$, and $L_C > L$  for $L>1$. This implies that  $L_C \ge L$ for any value of $m$.
$\square$}

\begin{table*}[!ht]
\centering
\caption{Crossbar node attain distinct conductance values for each of $M$ memristor having $L$ conductance state. }
\label{table3}
\begin{tabular}{p{2cm}cp{8cm}c}\\\hline
Number of memristors ($M$)& Number of levels ($L$)&  Unique {conductance} ($L_C$) &Sequence type\\\hline
1&$\{1,2,3,4,5,6,7,8,9,10,11,12..\}$ & $\{1,2,3,4,5,6,7,8,9,10,11,12..\}$&1-simplicial numbers\\
2&$\{1,2,3,4,5,6,7,8,9,10,11,12..\}$ & $\{1,3,6,10,15,21,28,36,45,55,66,78,..\}$&2-simplicial numbers\\
3&$\{1,2,3,4,5,6,7,8,9,10,11,12..\}$ & $\{1,4,10,20,35,56,84,120,165,220,286,364,..\}$&3-simplicial numbers\\
4&$\{1,2,3,4,5,6,7,8,9,10,11,12..\}$ & $\{1,5,15,35,70,126,210,330,495,715,1001,1365,..\}$&4-simplicial numbers\\
5&$\{1,2,3,4,5,6,7,8,9,10,11,12..\}$ & $\{1,6,21,56,126,252,462,792,1287,2002,3003,4368,..\}$&5-simplicial numbers\\
6&$\{1,2,3,4,5,6,7,8,9,10,11,12..\}$ & $\{1,7,28,84,210,462,924,1716,3003,5005,8008,12376,..\}$&6-simplicial numbers\\
7&$\{1,2,3,4,5,6,7,8,9,10,11,12..\}$ & $\{1,8,36,120,330,792,1716,3432,6435,11440,19448,31824,..\}$&7-simplicial numbers\\
8&$\{1,2,3,4,5,6,7,8,9,10,11,12..\}$ & $\{1,9,45,165,495,1287,3003,6435,12870,24310,43758,75582,..\}$&8-simplicial numbers\\\hline

\end{tabular}
\end{table*}

\begin{table*}[!ht]
\centering
\caption{Example of unique conductance values obtained from the crossbar nodes with combination of 1, 2 and 3 memristors each having $L=\{1,2,3\}$ conductance levels. }
\label{table:toyexample}
\begin{tabular}{p{0.5cm}|p{2cm}|p{1.1cm}|p{4cm}|p{1cm}|p{6cm}|p{1cm}}\hline
$L$&\multicolumn{2}{|c}{1-M}& \multicolumn{2}{|c}{2-M} & \multicolumn{2}{|c}{3-M}\\\cline{2-7}
&Conductance value& $L_C=L$&Conductance value& $L_C$&Conductance value& $L_C$\\\hline
1& $g_1$ &1&$2g_1$&1& $3g_1$&1\\
2& $g_1,g_2$ &2 &$2g_1,2g_2,(g_1+g_2)$&3& $3g_1,3g_2,g_1+2g_2,g_2+2g_1$&4\\
3& $g_1,g_2,g_3$ &3 &$2g_1,2g_2,2g_3,g_1+g_2,g_1+g_3,g_2+g_3$&6&$3g_1,3g_2,3g_3,2g_1+g_2,2g_1+g_3,2g_2+g_1,2g_2+g_3,2g_3+g_1,2g_3+g_2,g_1+g_2+g_3$&10\\\hline
\end{tabular}
\end{table*}

\begin{figure}[!ht]
    \centering
    \includegraphics[width=90mm]{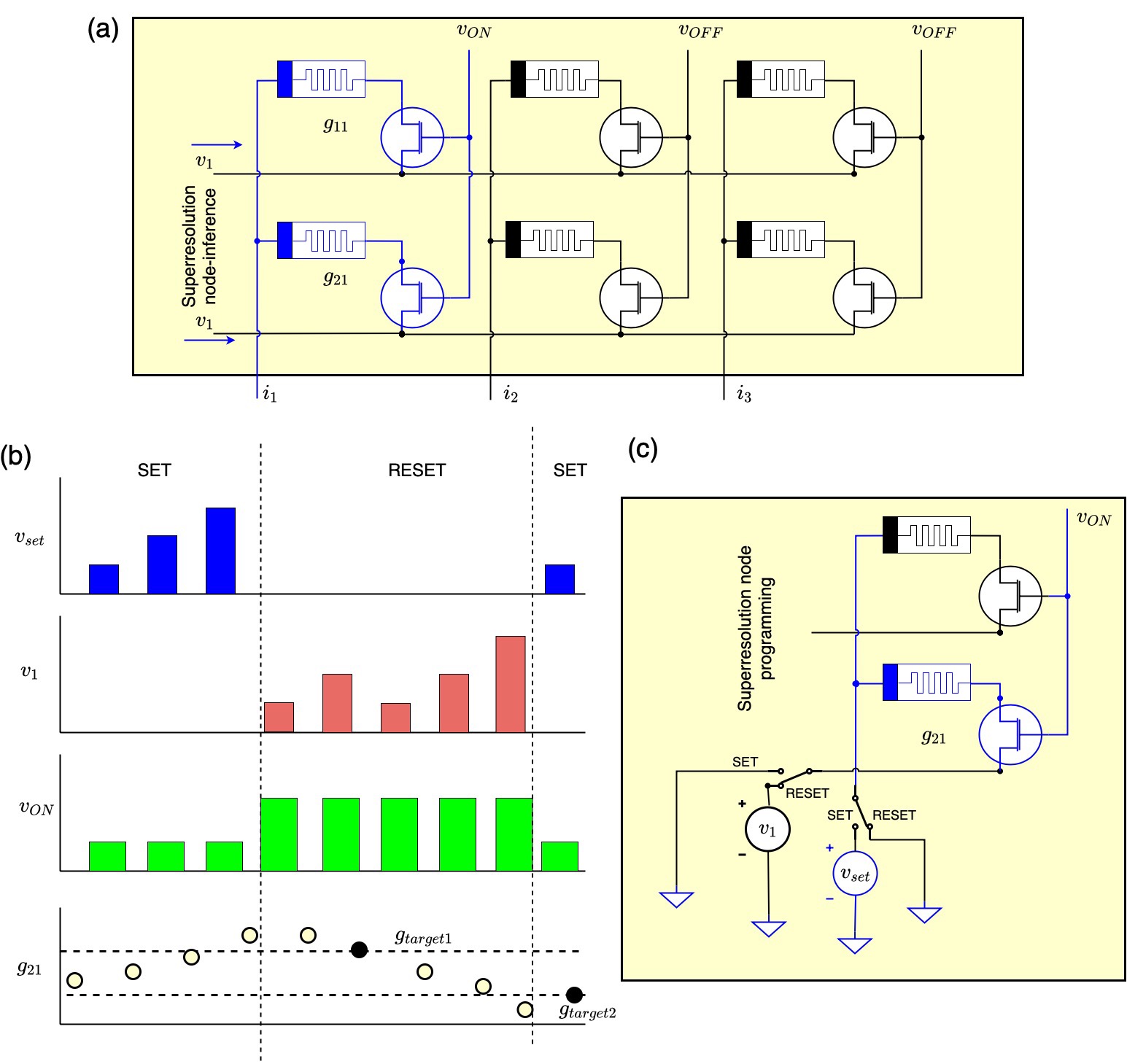}
    \caption{{Programming of the superresolution node. (a) Inference schematic, (b) SET and RESET sequence for programming the set conductance to a desired value, and (c) Programming schematic.} }
    \label{newfig}
\end{figure}

\begin{figure*}[!ht]
    \centering
    \includegraphics[width=200mm]{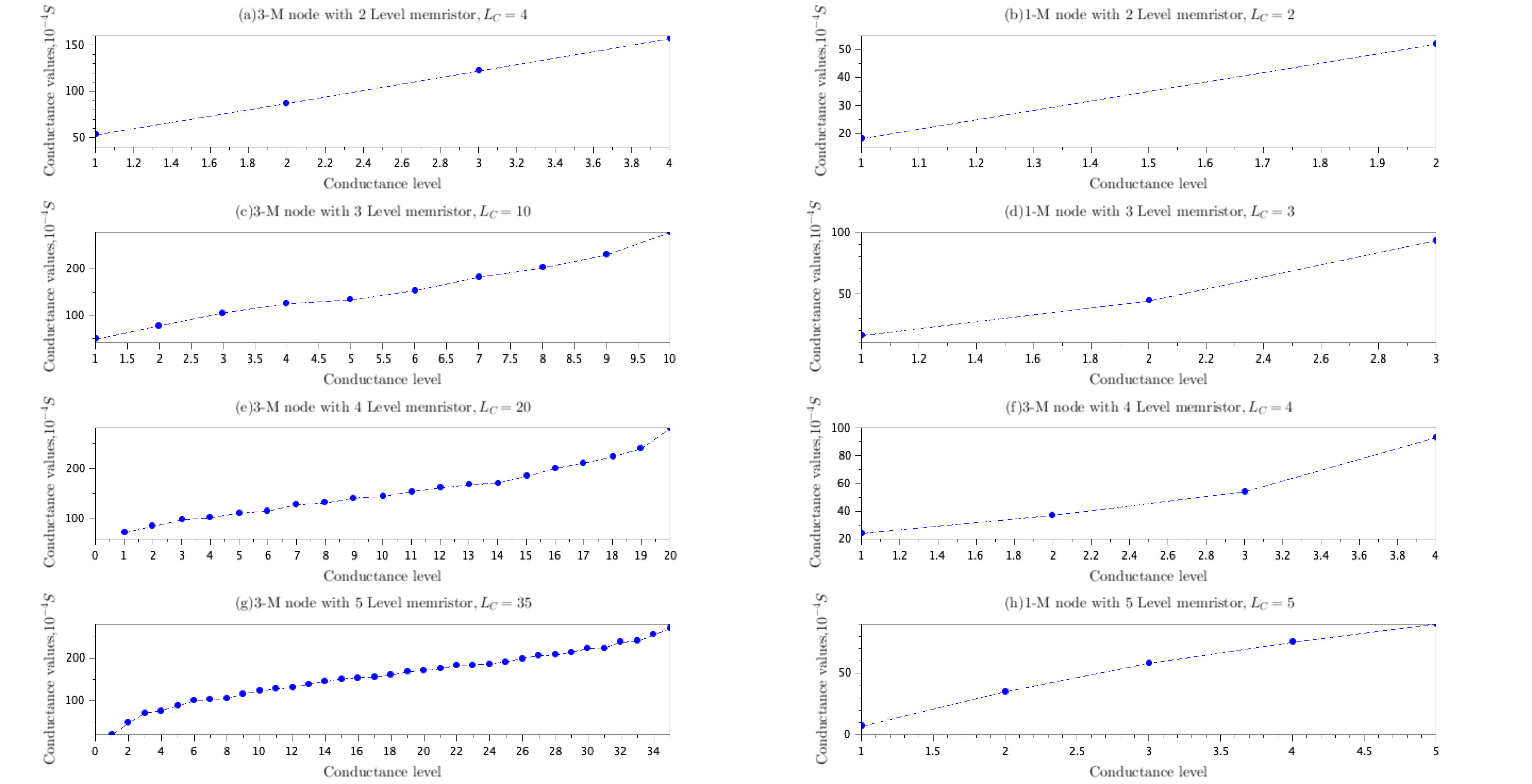}
    \caption{Example comparison of crossbar node with single memristor (1-M) with three memristor (3-M). The sub-figures (b,d,f,h) are for 1-M node with unique conductances $L_C$ of 2, 3, 4 and 5, respectively. The sub-figures (a,c,e,g) shows unique conductance for 3-M nodes that are {built} using corresponding 1-M nodes in (b,d,f,h). }
    \label{fig:toyexample}
\end{figure*}

\
\deleted{Corollary to Theorem 2, is that for \emph{\textbf{\(m\ge2\)}} any increase in \emph{\textbf{\(m\)}} results in \(L_C>L\).}

In such a  generalization, both the number of conductance levels and memristors can be increased. Currently, there are memristors that can have more than 100 conductance levels~\cite{Mannion_2020}. However, most of the memristor devices have only very few stable conductance levels that are easily programmed under the influence of integrated circuit parasitics and aging. To build analog crossbar arrays that can map weights at high accuracy and precision, it must have a higher number of levels.

The proposed approach, when extended to include the impact of a higher number of conductance levels per device, implies a higher number of possible conductance per node.
Table {\ref{table3}} shows the generalization of this scenario. It is observed that the unique conductance per node for a given number of memristors follows \(r\)-simplicial numbers. With memristors having even 3 stable levels, increasing the number of memristors per node can lead to a large
number of conductance levels per node {as seen in Tables {\ref{table3}} and  {\ref{table:toyexample}}}. As the number of levels per memristor increases, fewer memristors are required to achieve a high number of levels per node. For example, a node made up of 8 memristors, each having 8 levels, will result in 6435 unique conductance per node.

\subsection{Illustrative Example} 

Table \ref{table:toyexample} shows the conductance calculations for crossbar nodes made up of one memristor (1-M), two memristors (2-M), and three memristors (3-M)\deleted{, and each}\added{. Each} memristor has $L$ conductance levels. {The $L_C$ for 1-M follows the set \{1,2,3..\}, 2-M follows the set \{1,3,6..\} and 3-M follows the set \{1,4,10..\}. Thus, the $m$-M node follows the simplicial number sequence $\{ 1, \frac{m^{(1)}}{1!}, \frac{m^{(2)}}{2!},...\}$, where $m$ is the number of memristors per node. Figure \ref{newfig} shows the programming scheme and using the superresolution nodes in the inference stage. The programming of the memristors in each node is achieved by applying the SET and RESET cycles of pulses. For the programming stage, the desired conductance states $g_{target1}$ and $g_{target2}$ for the 1T1M memristor node can be obtained by repeated SET and RESET cycles, as shown in Fig. \ref{newfig}(b). The switch over between the SET and RESET operations use the switching logic shown in Fig. \ref{newfig}(c). {The voltage pulses induce low potential difference during read (or inference or verify operation), high positive potential difference during  program operations, and negative potential difference during the erase operation across the memristor.} During the inference, the crossbar {does} not need any modification, and superresolution nodes can be formed by feeding the same voltage $v_1$ to $g_{11}$ and $g_{21}$ 1T1R nodes to form $1\times 2$ sized superresolution node, as shown in Fig. \ref{newfig}(a).}

Figure \ref{fig:toyexample} shows the example simulation of conductance level distribution resulting from a 3-M node compared with a 1-M node. The 3-M nodes built with memristors having 2, 3, 4, and 5 unique conductance levels can result in 4, 10, 20, and 35 unique conductance levels, respectively.  In this example, the memristors' conductance levels are randomly assigned to a stable value that the individual memristor can be programmed. The resulting 3-M node shows a higher number of stabilized and unique conductance values. Therefore, the 3-M based crossbar can have a higher weight resolution in a neural network than 1-M nodes. The conductance drift due to aging in 1-M nodes can result in loss of conductance levels that cannot be recovered quickly. {Even if one of the memristor loss levels in a 3-M, the available conductance levels ($L$) in the remaining two memristors can still result in large values of $L_C$ for the superresolution nodes.} The example also shows that individual memristors in a 3-M node can easily program the states to a few stable states distributed in any fashion. The resulting conductance space will be large, allowing for more considerable weight precision.

\paragraph*{Programming sequence} An example sequence to program the  conductance values of the 3-M node, using three memristors each having four levels \deleted{are}\added{is} shown in Table \ref{mapping}. The node conductance $g_n$ can be mapped to the weights in the neural network application or general in-memory computing applications using a lookup approach. Suppose \deleted{the}\added{a} weight of 0.3 is mapped to the conductance of 40 $\mu S$, the three memristors in the 3-M node are required to be programmed to the combination without following the sequenced order 15, 15, and 10 $\mu S$ values. The intermediate weight values such as 0.33 are rounded off to 0.3 by the nearest distance search and subsequently programmed to the value from the lookup table, in Table \ref{mapping}. In contrast, a 1-M node with four levels would have resulted in a significant loss of weight precision as many intermediate levels are missed out.   Through this approach, the complexity of programming crossbar nodes to obtain high-resolution weights substantially reduces, as fewer levels are often easier to program and are more reliable to achieve. In addition, the super-resolution nodes are formed by a parallel or series combination of memristors, depending on the ease of programming conductance or resistance of individual memristors, respectively. More details on programming memristor based on the resistance or conductance are included in the Supplementary material. 

\begin{table}[]
\centering
\caption{Programming sequence for memristor with 4 levels (10,15,29,1000)$\mu S$ for a 3-M node.}
\label{mapping}
\begin{tabular}{cccc}\hline
\multicolumn{3}{l}{3-M node conductance($\mu S$)} & \multicolumn{1}{l}{3-M node conductance($\mu S$)}  \\\hline
$g_1$&$g_2$ &$g_3$                          & \multicolumn{1}{c}{$g_n$ }                       \\\hline
10&10&10     & 30      \\
15&10&10     & 35      \\
15&15&10     & 40        \\
15&15&15     & 45       \\
29&10&10     & 49        \\
29&15&10     & 54        \\
29&15&15     & 59       \\
29&29&10     & 69      \\
29&29&15     & 74       \\
29&29&29     & 88        \\
1000&10&10   & 1020    \\
1000&15&10  & 1025    \\
1000&15&15  & 1030   \\
1000&29&10  & 1039    \\
1000&29&15  & 1044   \\
1000&29&29  & 1059   \\
1000&1000&10 & 2010   \\
1000&1000&15 & 2015   \\
1000&1000&29 & 2029  \\
1000&1000&1000 & 3000    \\\hline                             
\end{tabular}
\end{table}

\begin{figure}[!ht]
    \centering
    \includegraphics[width=60mm]{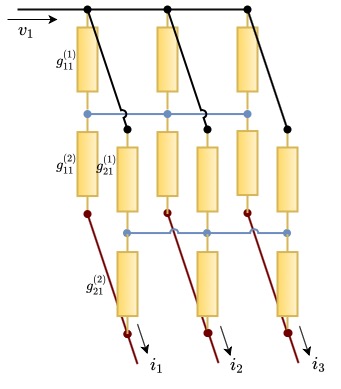}
    \caption{{Extension of the conductance range using 3D crossbar configuration. Note that each memristor comes with a selector switch to program and select the column.}}
    \label{fig:3d}
\end{figure}

\paragraph*{Case of 3D crossbar} As an extension, the \added{resistance} range \deleted{of the resistance values} can be further increased \deleted{by} using a 3D crossbar array. { In the example shown in Fig. \ref{fig:3d}, the top and bottom layer conductances are denoted as $g_{ij}^{(1)}$, and $g_{ij}^{(2)}$, respectively. Because each node is a parallel combination of two memristors connected in series across two layers, current $i_1 = v_1*(g_{11}^{(1)} || g_{11}^{(2)})+v_1*(g_{21}^{(1)} || g_{21}^{(2)})$. Each memristor node has a selector switch. When current $i_1$ is read, the memristor nodes in columns $i_2$ and $i_3$ are kept to OFF state. Hence, $g_{11}^{(1)}$ will be in series to $g_{11}^{(2)}$}
{ In general, the read out currents from any one column is given as, $i_k=\sum v_k g_{kl} $, with the equivalent node conductance as follows:
\begin{equation}
    g_{kl}=\left[\sum \frac{1}{g_{ik}^{(1)}}\right]^{-1}   + \left[ \sum \frac{1}{g_{ik}^{(2)}}\right]^{-1}.
\end{equation}
 Such 3D nodes can further extend the number of $L_C$ levels without significantly increasing the area on-chip. The different configurations and possibilities with the 3D crossbar array are a topic for future study.}

\deleted{In the crossbar illustrations in Fig. 1, two inputs are shown, while in Figs. 2,4 one input each is shown, while it should be implied that several such input blocks when connected column-wise form a crossbar for analog computing.}

\section{Results and Discussion}

{\label{460767}}

\begin{figure}[!ht]
    \centering
    \includegraphics[width=80mm]{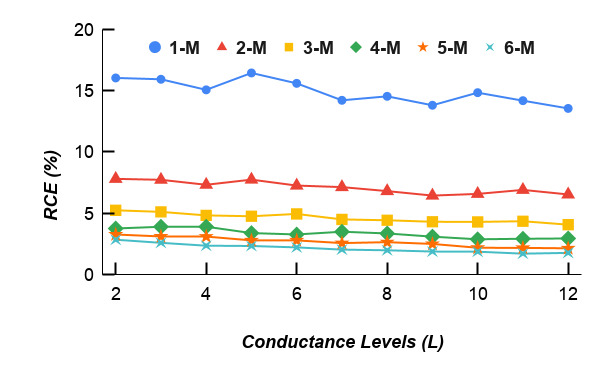}
    \caption{Relative output current error percentages in a crossbar having 10x10 nodes, with $R_{OFF}/R_{ON}=100k\Omega/1k\Omega$.}
    \label{tabcurrentf}
\end{figure}

The crossbars are used in a range of applications, mainly, in building
neural network accelerators. To check the impact of the number of
conductance per node on the overall performance of the crossbar, we
take a crossbar array with 10$\times$10 nodes. The conductance programming is performed using an adaptive program and verify approach \cite{aaidana2020}. The output currents are read
from the columns for the different memristors and conductance levels per node. {Throughout the paper, unless mentioned otherwise, random current read errors are added as a Gaussian distribution with mean zero and  standard deviation as 10\% of ideal read currents   to account for the non-idealities of the sensing circuits connected to the crossbar.}
Figure \ref{tabcurrentf} shows the relative output
current error (RCE) in percentage, calculated as~\(\frac{100\times\left|I_{o-ideal}-I_{o-real}\right|}{I_{o-ideal}}\),
where~\(I_{o-ideal}\) is the output current obtained when nodes have
infinite {number of} conductance levels, and~~\(I_{o-real}\) is the output
current obtained with a limited number of conductance levels.

\paragraph{Precision} The precision of the weights mapping to crossbar nodes is an important factor that determines the accuracy of analog dot product computations. {It can be observed from~Fig.~{\ref{tabcurrentf}} that
as the number of memristor per node increases, the current
error decreases. The increase in $L$ along with increase in $m$ results in a large increase in $L_C$, as shown in Table III. As $L_C$ increases, i.e. for larger value of $m$ and $L$, the current error tends to zero.} \deleted{ number of unique conductance per node increases the current
error decreases. Similarly,~the current errors also decrease as the
number of memristors per node increases.~ Table
~{\ref{tabcurrent}} shows the crossbar node with 6
memristors, with each memristor having 8 conductance levels resulting in
near zero current errors.} {The memristors in the superresolution crossbar are connected in parallel and the current readout from the nodes will be the weighted average of the input voltage, i.e. $i=\sum_{i=1}^m g_i v_{in}$, where $v_{in}$ input voltage to a node with $m$ memristors each having its own conductance $g_i$. Suppose there is a variation in one of the conductance values say in the $m$th memristor i.e. $g_m=g_i\pm \delta$, with $\delta$ being the offset from the true value, \deleted{then} then we can write, $i=\left[\sum_{i=1}^{m-1} g_i v_{in}\right]\pm g_m v_{in}$ or $i=\sum_{i=1}^{m} g_i v_{in}\pm \delta v_{in}$. Thus the variations, i.e. $\delta$, will have less impact on the overall current $i$ as the value of $m$ increases.}

\begin{figure*}[ht]
  \centering
  \subfigure[1-M]{\includegraphics[width=50mm]{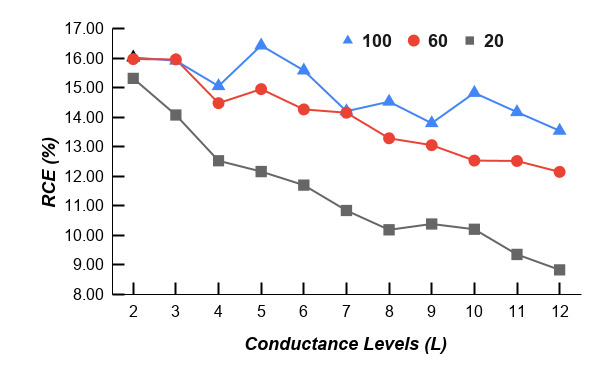}}
  \subfigure[2-M]{\includegraphics[width=50mm]{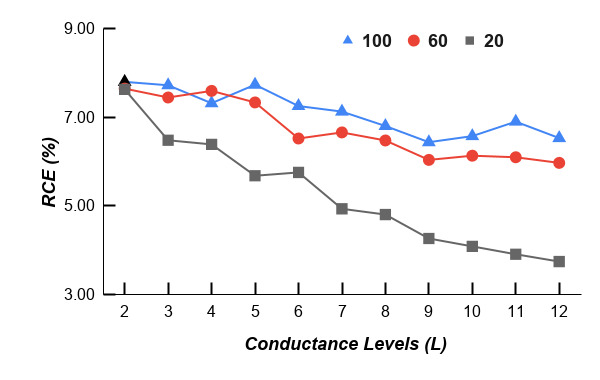}}
   \subfigure[3-M]{\includegraphics[width=50mm]{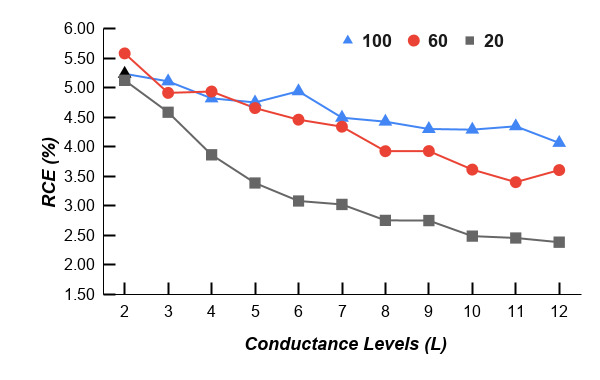}}
   \subfigure[4-M]{\includegraphics[width=50mm]{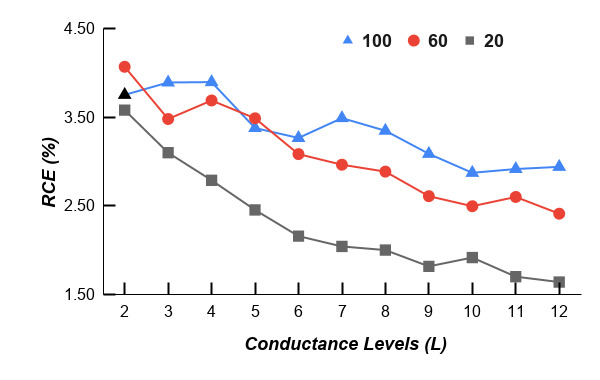}} 
   \subfigure[5-M]{\includegraphics[width=50mm]{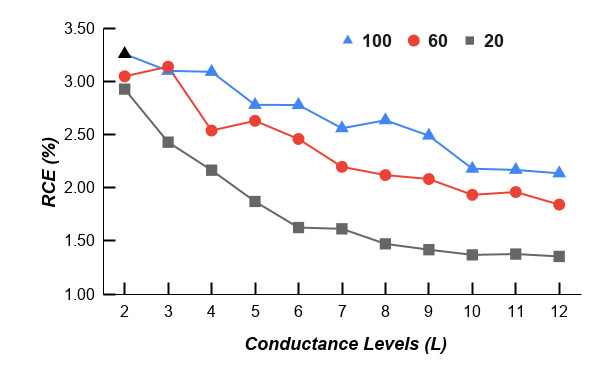}}
   \subfigure[6-M]{\includegraphics[width=50mm]{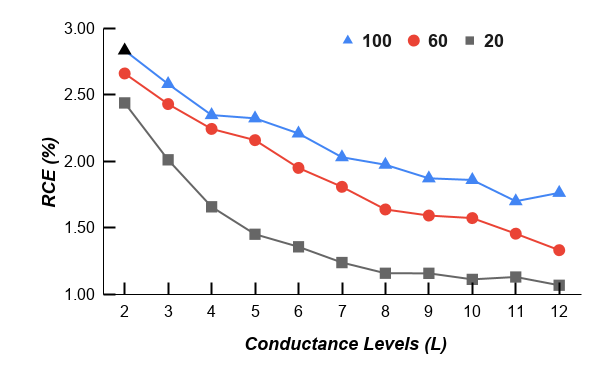}}
\caption{{Impact of the change in $R_{OFF}/R_{ON}$ ratios (i.e. 100, 80,  60, 40, 20, 10, 5) relative to the change in node size (i.e. 1-M, 2-M, 3-M, 4-M, 5-M, 6-M) and conductance levels. }}
\label{fig:fig4}
\end{figure*}

{Figure \ref{fig:fig4} shows the average relative current errors (RCE), with changes in $R_{OFF}/R_{ON}$ ratio of memristor relative to the node size.  The larger the $R_{OFF}/R_{ON}$ ratio, the larger the RCE values irrespective of the node size.  However, as the number of memristors increases in the node, the RCE values drastically reduce irrespective of the changes in $R_{OFF}/R_{ON}$ ratio, indicating the crossbar superresolution's usefulness. }

\begin{figure}[!ht]
    \centering
   \includegraphics[width=90mm]{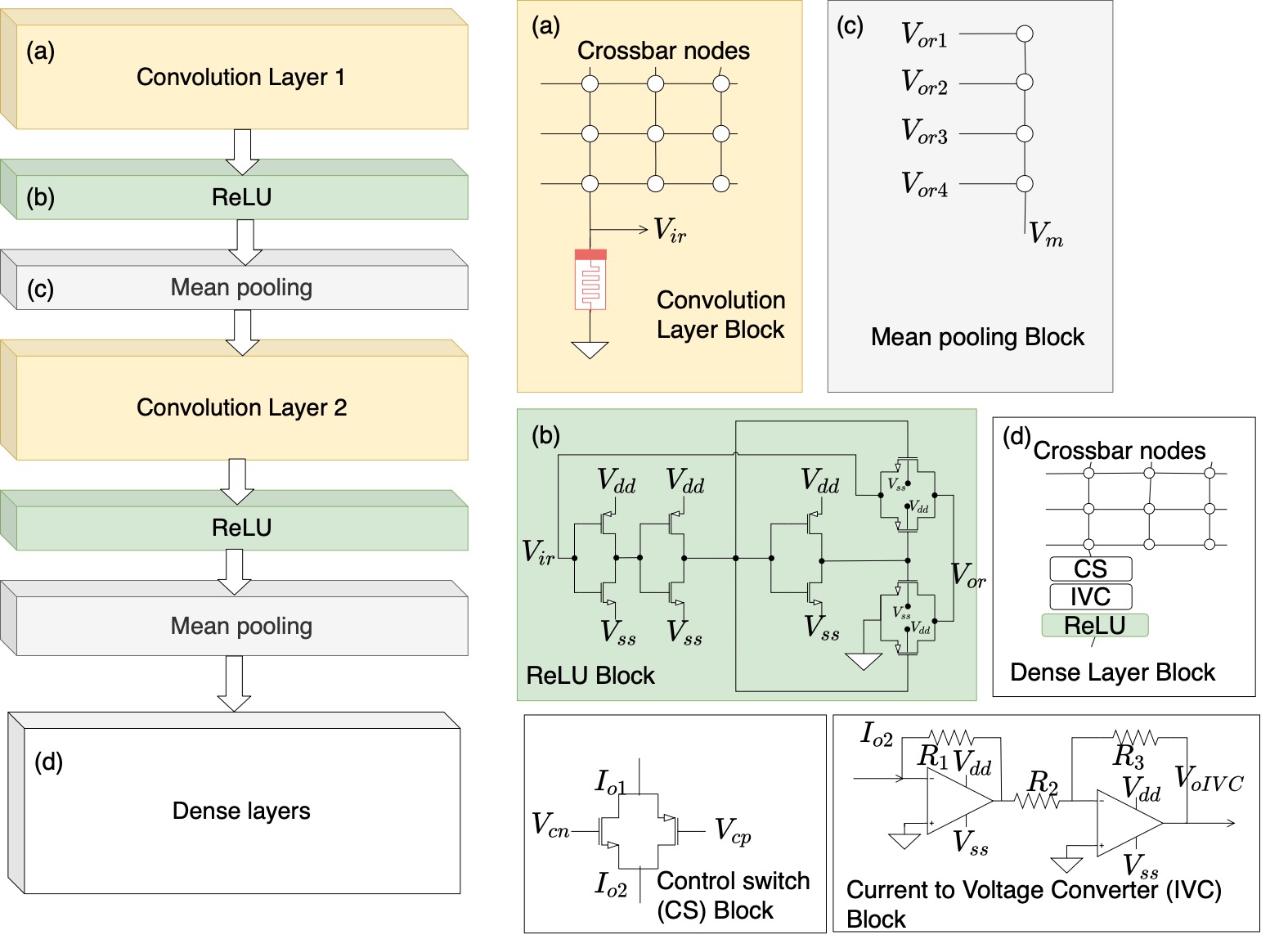}
    \caption{{Various circuit block in the CNN. (a) shows the circuit block for convolution layer, (b) ReLU circuit, (c) mean pooling block, and 
    (d) dense layer block.}}
    \label{figcircuit}
\end{figure}

\paragraph{Area and power} {\deleted{At the maximum each}\added{Each} memristor \deleted{will have}\added{has} one selector transistor switch. Irrespective of the node size, i.e.,  nodes with $m$ memristor $m$T-$m$M, the interface circuits required at the output of the \deleted{crossbar for an} analog neural network \added{crossbar }remain the same. {If we assume the area occupied by interface circuits connected to each of the crossbar column to be $A_i$,  and the area of each \deleted{node with} 1T1M node to be $A_c$, then the minimum total area for implementing one layer of a neural network will be $kA_i+nmkA_c$, where $k$ is the number of columns in the crossbar, $n$ is the number of nodes in a column, and $m$ is the number of memristors per node.} In analog neural network implementations \cite{krestinskaya2020memristive}, $A_i>>A_c$, for example, in a 180 nm CMOS technology\footnote{The 180 nm CMOS node used is for demonstrative purpose only. It is possible to use lower node sizes such as 22 nm or lower FinFETs \cite{jain201913} node sizes for reducing the area and power requirements in practical applications. {For memristors, we used Knowm MSS (Multi-Stable Switch) memristor model and WOx device parameters \cite{knowmorg}.}}, the interface circuits to each column of the crossbar consists of I-V converter block, Opamp buffer, and ReLU block, occupying areas of 96.714 $\mu$m$^2$, 44.712$ \mu$m$^2$, and 23.659$ \mu$m$^2$, respectively, {as shown in Fig. \ref{figcircuit}}\footnote{{Here, the ReLU PMOS has W/L=0.36/0.18, and ReLU NMOS has W/L=1/0.18. Switch transistor sizes are 5/0.18 (PMOS), and 20/0.18(NMOS), and feedback resistors $R_1=R_2=R_3=1k\Omega$}}. The single 1T1M memristor node occupies an area of  0.14$\mu$m$^2$. {For a $10 \times 10$ super-resolution crossbar, with 8T8M nodes, the minimum areas are $kA_i=1650.85 \mu m^2$, and $nmkA_c=11.2 \mu m^2$, where $n=10$, $m=8$, $k=10$. Alternatively, if we use a classical crossbar, 1T1M nodes spread across 8 columns, the equivalent crossbar will have a size of $10 \times 80$, with minimum areas of $kA_i=13206.8 \mu m^2$, and $nmkA_c=11.2 \mu m^2$, where $n=10$, $m=1$, and $k=80$.}  In addition, as the lower and upper limits of resistance values in the crossbar node (i.e., $R_{ON}$ and $R_{OFF}$ limits) does not change, and reverse leakage currents in the transistors are relatively lower than the forward currents in the memristor, the net maximum current output range out of the crossbar columns do not change. {The average power dissipation per convolution filter when using a  ReLU block is 40.16 mW \footnote{This can be further scaled down to 1.63 mW if using a lower technology such as 45 nm.}.  The delay associated with this block is 1.28 ns. For the dense layer crossbar implemented with ReLu, the power comes to 441.9 mW, with a propagation delay of 1.45 ns.} The overall maximum power consumption is mainly impacted by the non-idealities of the transistor, which can be handled by using FinFETs. Further, even if there is an increase in memristors per node and size of the crossbar, the readout circuits connected to the crossbar columns do not need to be changed if the output current range is kept the same by proportionally scaling the conductance levels. }

\paragraph*{Case of scaling the crossbar} As the crossbar array size increases, the sneak path currents and leakage currents in the switches that add to the current errors also increase. \deleted{To avoid this}\added{Therefore}, the large crossbar arrays \deleted{should be}\added{are} split into smaller arrays, known as the modular crossbar array or tiling of crossbar arrays. In analog crossbars tiling, the input space is split across several crossbars, and their column-wise output is summed up. The area overhead for this is a summing amplifier, which adds another 4.712$\mu m^2$ per column while allowing for very large crossbar implementations.

\begin{figure*}
    \centering
     \subfigure[1-M]{\includegraphics[width=50mm]{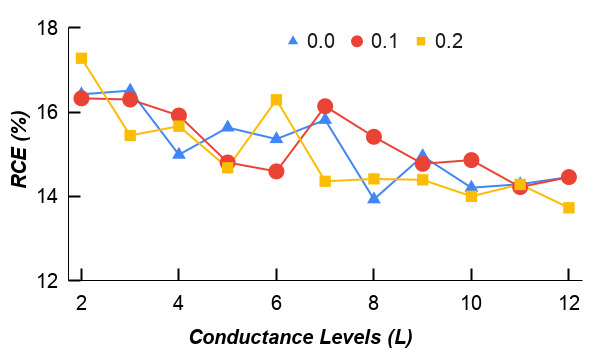}}
     \subfigure[2-M]{\includegraphics[width=50mm]{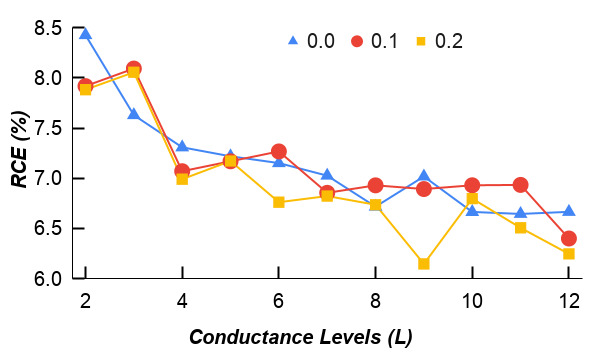}}
     \subfigure[3-M]{\includegraphics[width=50mm]{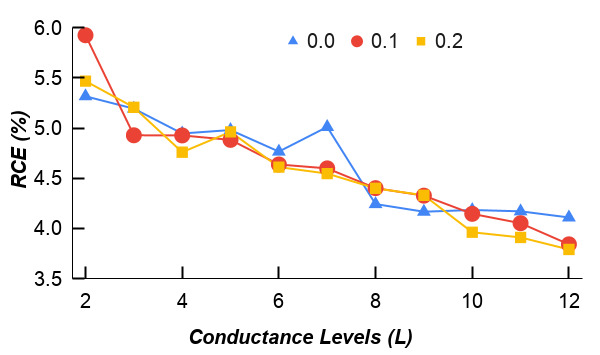}}
     \subfigure[4-M]{\includegraphics[width=50mm]{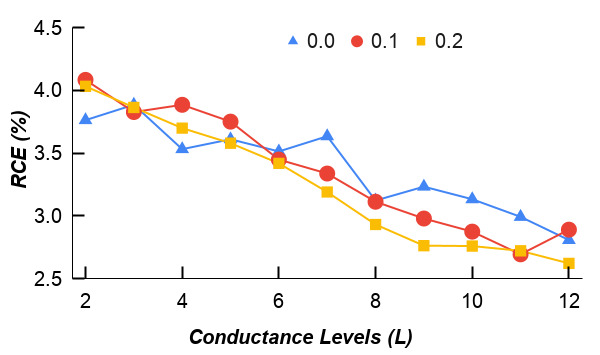}}
     \subfigure[5-M]{\includegraphics[width=50mm]{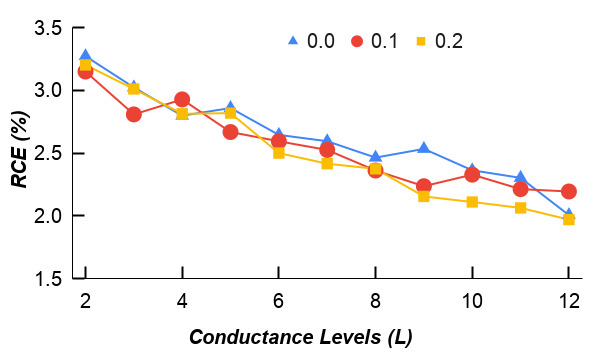}}
     \subfigure[6-M]{\includegraphics[width=50mm]{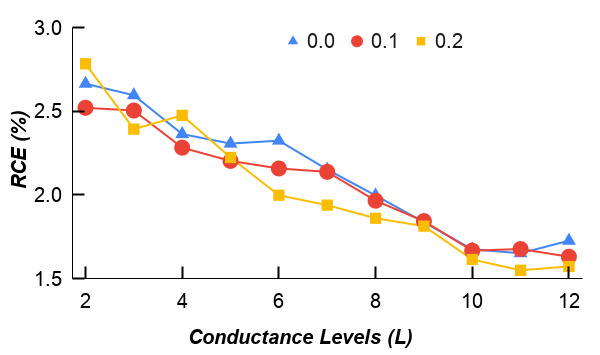}}
    \caption{{Impact of type 3 aging (i.e. with $R_{OFF}$ decreasing and $R_{ON}$ increasing) on the accuracy of the crossbar output currents, for different node size, and with various aging ratios. } }
    \label{figage}
\end{figure*}

\paragraph{Aging} 
{Three types of aging are observed in memristors (for details, see supplementary material). Higher the aging, the more the number of conductance levels that are lost. In our study, we select type 3 aging, which increases the value of $R_{ON}$ and decreases $R_{OFF}$ with repeated stress on the device. Setting the memristor  $R_{OFF}/R_{ON}$ ratio as 100, the effect of aging is observed for different node sizes. The nodes with single memristors, irrespective of the number of levels, tend to have high error levels. The aging ratios shown in the legend of Fig. \ref{figage} indicate the ratio of change in  $R_{OFF}$  and $R_{ON}$ with respect to its original value. For example, the aging ratio of 0.1 means the new resistance to be $0.9R_{OFF}$ and $1.1R_{ON}$. The states are reprogrammed in these experiments; hence, attempts are made to recover the number of conductance states that disappear with aging. The nodes with more than three memristors show less than 10\% relative current errors under a high aging ratio of 0.7, irrespective of the number of conductance levels. In all cases, the errors are higher for aging when the number of conductance levels is low. The increase in the number of levels reduces the relative current errors.  }

The memristors, like any other semiconductor device, get affected by
non-idealities and aging. As the devices are used continuously for
extended periods, the conductance states of the memristors gradually
become harder to program.~ When the device has a large number of
conductance levels over a period, the number of conductance
levels decreases. \deleted{The devices having}\added{Having} a larger number of levels would
\deleted{become less reliable}\added{reduce reliability} as even minor variations in the conductance \deleted{will
result in a reduction of}\added{reduce the} conductance levels. The crossbar node (165
unique conductance) built with eight memristors, each having four
conductance levels will be more robust than a crossbar node~(165 unique
conductance) with three memristors each having nine conductance levels.

\begin{figure*}
    \centering
    \subfigure[1-M]{\includegraphics[width=50mm]{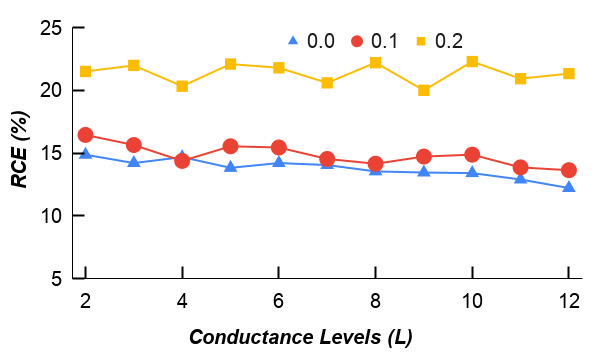}}
     \subfigure[2-M]{\includegraphics[width=50mm]{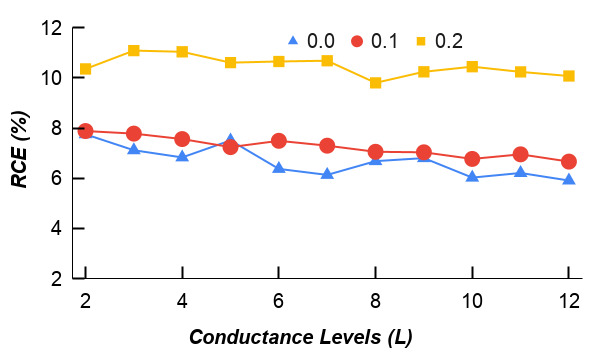}}
     \subfigure[3-M]{\includegraphics[width=50mm]{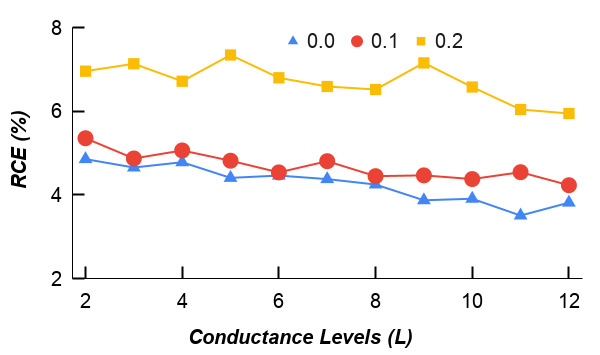}}
     \subfigure[4-M]{\includegraphics[width=50mm]{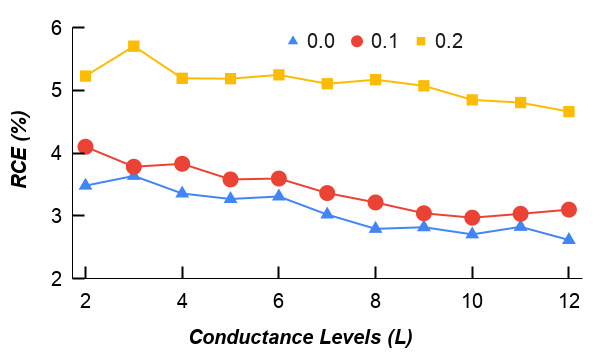}}
     \subfigure[5-M]{\includegraphics[width=50mm]{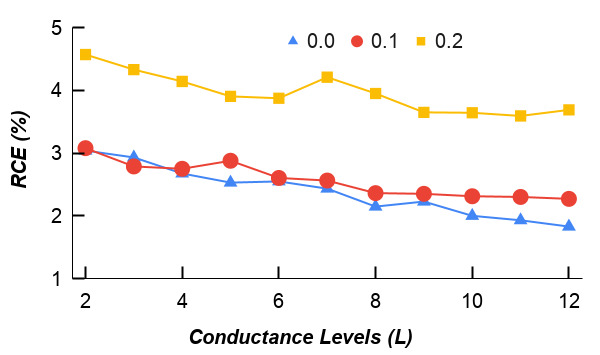}}
     \subfigure[6-M]{\includegraphics[width=50mm]{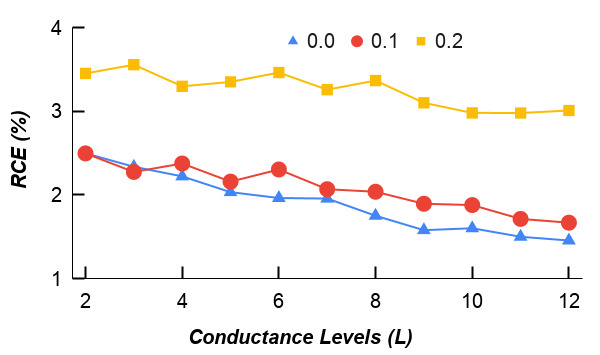}}
    
    \caption{{Impact of input signal noise on superresolution crossbar. The performance of the crossbar is compared for additive Gaussian noise with different variance values.}}
    \label{noise}
\end{figure*}

\paragraph{Signal noise} {The input signals to the crossbar are prone to noise. This noise can result from sensors or \deleted{due to} signal integrity issues. To test the impact of signal noise on the superresolution crossbar, we apply additive Gaussian noise with variances, as shown in Fig. \ref{noise}. \deleted{Here, $v_i$ is the input signal without noise, and $\Delta$ is a random Gaussian noise generator with mean value zero and variance value one.} The RCE values indicate that the superresolution nodes do not impact the relative accuracy of the current outputs relative to the node with ideal weights.}

\paragraph{Effects of wire resistance, read instabilities and $R_{ON},R_{OFF}$ variability }

{
To analyse the impact of the crossbar variability, we perform boundary analysis \deleted{by take into account}\added{considering the} combined effect of wire-resistance,  $R_{ON}$, $R_{OFF}$ changes in the crossbar. At each node, the maximum variation is assumed to be 2.5 $\pm$ 0.25 $\Omega$ for wire resistance, boundary resistances change as $R_{ON}\pm 0.2R_{ON}$ and $R_{OFF}\pm 0.2R_{OFF}$. Table \ref{tablewire} shows that when there is wire resistance and changes in boundary resistances, the impact is more evident in a single memristor node and less prominent in super-resolution nodes. Here, N and Y indicate the cases  without and with wire resistance, respectively. In both cases, the boundary resistances changes by a maximum 20\% standard deviation from the mean $R_{ON}$ and $R_{OFF}$. Read instability \cite{wiefels2020hrs,puglisi2018random} can also occur in a crossbar, with the current readouts varying from their mean value. Inclusion of a high read instability, of 10\%, on the crossbar readout currents, are shown in the column labelled R in Table \ref{tablewire}. Even in this case, the use of multiple memristors per node show far more robustness than single memristor per node crossbar.}

\begin{table}[!ht]
\caption{RCE (\%) values showing impact of wire resistance and read errors with $R_{ON}\pm 0.2R_{ON}$ and $R_{OFF}\pm 0.2R_{OFF}$. }
\label{tablewire}
\scalebox{.6}{
{
\begin{tabular}{|r|r|r|r|r|r|r|r|r|r|r|r|r|r|r|r|r|r|r|}\hline
&\multicolumn{3}{l|}{\textbf{1-M node}} & \multicolumn{3}{l|}{\textbf{2-M node}}&\multicolumn{3}{l|}{\textbf{3-M node}}&\multicolumn{3}{l|}{\textbf{4-M node}}&\multicolumn{3}{l|}{\textbf{5-M node}}&\multicolumn{3}{l|}{\textbf{6-M node}}\\\cline{2-19}
$L$& N & Y & R & N & Y & R& N & Y & R& N & Y & R& N & Y & R& N & Y& R \\\hline
2& 14.7 & 15.5 & 15.7 & 7.8 & 7.8 & 8   & 5.3 & 5.3 & 5.3 & 3.7 & 3.7 & 3.9 & 2.9 & 3   & 3.4 & 2.6 & 2.6 & 2.7 \\
3&7.9  & 8.2  & 10   & 3.8 & 3.8 & 4.7 & 2.5 & 2.4 & 3   & 1.8 & 1.9 & 2.5 & 1.5 & 1.4 & 1.9 & 1.2 & 1.3 & 1.6 \\
4&5.3  & 5.3  & 8.1  & 2.5 & 2.5 & 4   & 1.7 & 1.7 & 2.8 & 1.2 & 1.2 & 2   & 1   & 1   & 1.5 & 0.9 & 0.9 & 1.4 \\
5&3.6  & 3.8  & 7.2  & 1.8 & 1.8 & 3.5 & 1.3 & 1.2 & 2.6 & 0.9 & 1   & 1.7 & 0.8 & 0.7 & 1.4 & 0.6 & 0.6 & 1.1 \\
6&2.9  & 2.9  & 6.8  & 1.4 & 1.5 & 3.4 & 1   & 1   & 2.2 & 0.8 & 0.7 & 1.6 & 0.6 & 0.6 & 1.4 & 0.5 & 0.5 & 1.1 \\
7&2.5  & 2.5  & 6.8  & 1.3 & 1.3 & 3.5 & 0.8 & 0.8 & 2.2 & 0.6 & 0.6 & 1.7 & 0.5 & 0.5 & 1.3 & 0.4 & 0.4 & 1.1 \\
8&2.0    & 2.2  & 6.6  & 1   & 1.1 & 3.5 & 0.8 & 0.7 & 2.2 & 0.5 & 0.5 & 1.7 & 0.4 & 0.4 & 1.4 & 0.3 & 0.4 & 1.1 \\
9&2 .0   & 2.0  & 6.8  & 0.9 & 0.9 & 3.3 & 0.6 & 0.6 & 2.1 & 0.5 & 0.4 & 1.7 & 0.4 & 0.4 & 1.2 & 0.3 & 0.3 & 1   \\
10&1.7  & 1.8  & 6.4  & 0.8 & 0.8 & 3.4 & 0.6 & 0.5 & 2.1 & 0.4 & 0.4 & 1.7 & 0.3 & 0.3 & 1.2 & 0.3 & 0.3 & 1.1 \\
11&1.5  & 1.6  & 6.5  & 0.7 & 0.7 & 3.2 & 0.5 & 0.5 & 2.2 & 0.4 & 0.4 & 1.6 & 0.3 & 0.3 & 1.3 & 0.2 & 0.2 & 1   \\
12&1.3  & 1.4  & 6.4  & 0.7 & 0.7 & 3.2 & 0.5 & 0.4 & 2   & 0.3 & 0.3 & 1.6 & 0.3 & 0.3 & 1.3 & 0.2 & 0.2 & 1.1 \\\hline
\end{tabular}}}
\end{table}

\paragraph{Application in CNN}
{To perform the analysis and tests on the crossbar for an analog neural network accelerator application, we select the convolutional neural network and its analog hardware implementation outlined in \cite{krestinskaya2020memristive}.  The architecture consists of  two convolutional and two dense layers, with the circuits for implementing the dropout\footnote{Dropout is a technique to regularise neural network by dropping OFF neuronal nodes during the training, which avoids complex co-adaptations and over-fitting on training data.} and programming of memristor levels. The circuit design and parameter extraction for CNN circuit blocks in \cite{krestinskaya2020memristive} was done in SPICE, while architecture level simulations for calculating performance accuracy with and without variability were performed with customised Python programs. }

\begin{table}[!ht]
\centering
\caption{{Recognition accuracy when using a CNN accelerator with the proposed
memristor crossbar nodes. The tests are performed on CIFAR-10 dataset. The
ideal accuracy with infinite memristive states is 91.26\%. }}
\label{tabcnn}
\scalebox{.6}{
\begin{tabular}{|p{.5cm}|p{.5cm}|p{.5cm}|p{.5cm}|p{.5cm}|p{.5cm}|p{.5cm}|p{.5cm}|p{.5cm}|p{.5cm}|p{.5cm}|p{.5cm}|p{.5cm}|p{.5cm}|p{.5cm}|p{.5cm}|}
\hline
\multicolumn{2}{|l|}{\textbf{1-M node}}                    & \multicolumn{2}{l|}{\textbf{2-M
node}}                     & \multicolumn{2}{l|}{\textbf{3-M node}}     
               & \multicolumn{2}{l|}{\textbf{4-M node}}                 
   & \multicolumn{2}{l|}{\textbf{5-M node}}                     & \multicolumn{2}{l|}{\textbf{6-M
node}}                     & \multicolumn{2}{l|}{\textbf{7-M node}}     
               & \multicolumn{2}{l|}{\textbf{8-M node}}                 
   \\ \hline
\textbf{$L_C=L$} & \textbf{ACC (\%)} & \textbf{$L_C$} & \textbf{ACC (\%)} &
\textbf{$L_C$} & \textbf{ACC (\%)} & \textbf{$L_C$} & \textbf{ACC (\%)} &
\textbf{$L_C$} & \textbf{ACC (\%)} & \textbf{$L_C$} & \textbf{ACC (\%)} &
\textbf{$L_C$} & \textbf{ACC (\%)} & \textbf{$L_C$} & \textbf{ACC (\%)} \\
\hline
2                                                  & 82.43              
                         & 3                                            
     & 84.10                                        & 4                 
                                & 84.78                                 
      & 5                                                  & 85.19      
                                 & 6                                    
             & 86.66                                        & 7         
                                        & 89.07                         
              & 8                                                  & 90.07
                                       & 9                              
                   & 90.41                                        \\ \hline
3                                                  & 84.10              
                         & 6                                            
     & 86.66                                        & 10                
                                & 90.91                                 
      & 15                                                 & 91.14      
                                 & 21                                   
             & 91.20                                        & 28        
                                        & 91.18                         
              & \textbf{36}                & \textbf{91.26}       & 45  
                                              & 91.21                   
                    \\ \hline
4                                                  & 84.78              
                         & 10                                           
     & 90.91                                        & 20                
                                & 91.19                                 
      & 35                                                 & 91.24      
                                 & 56                                   
             & 91.28                                        & 84        
                                        & 91.22                         
              & 120                                                & 91.25
                                       & 165                            
                   & 91.28                                        \\ \hline
5                                                  & 85.19              
                         & 15                                           
     & 91.14                                        & 35                
                                & 91.24                                 
      & 70                                                 & 91.22      
                                 & 126                                  
             & 91.26                                        & 210       
                                        & 91.25                         
              & 330                                                & 91.26
                                       & 495                            
                   & 91.26                                        \\ \hline
6                                                  & 86.66              
                         & 21                                           
     & 91.20                                        & 56                
                                & 91.28                                 
      & 126                                                & 91.26      
                                 & 252                                  
             & 91.25                                        & 462       
                                        & 91.25                         
              & 792                                                & 91.26
                                       & 1287                           
                   & 91.26                                        \\ \hline
7                                                  & 89.07              
                         & 28                                           
     & 91.18                                        & 84                
                                & 91.22                                 
      & 210                                                & 91.25      
                                 & 462                                  
             & 91.25                                        & 924       
                                        & 91.26                         
              & 1716                                               & 91.26
                                       & 3003                           
                   & 91.26                                        \\ \hline
8                                                  & 90.07              
                         & \textbf{36}                & \textbf{91.26}  
    & 120                                                & 91.25        
                               & 330                                    
           & 91.26                                        & 792         
                                      & 91.26                           
            & 1716                                               & 91.26
                                       & 3432                           
                   & 91.26                                        & 6435
                                              & 91.26                   
                    \\ \hline
9                                                  & 90.41              
                         & 45                                           
     & 91.21                                        & 165               
                                & 91.28                                 
      & 495                                                & 91.26      
                                 & 1287                                 
             & 91.26                                        & 3003      
                                        & 91.26                         
              & 6435                                               & 91.26
                                       & 12870                          
                   & 91.26                                        \\ \hline
10                                                 & 90.91              
                         & 55                                           
     & 91.22                                        & 220               
                                & 91.27                                 
      & 715                                                & 91.26      
                                 & 2002                                 
             & 91.26                                        & 5005      
                                        & 91.26                         
              & 11440                                              & 91.26
                                       & 24310                          
                   & 91.26                                        \\ \hline
11                                                 & 90.89              
                         & 66                                           
     & 91.22                                        & 286               
                                & 91.27                                 
      & 1001                                               & 91.25      
                                 & 3003                                 
             & 91.26                                        & 8008      
                                        & 91.26                         
              & 19448                                              & 91.26
                                       & 43758                          
                   & 91.26                                        \\ \hline
12                                                 & 91.02              
                         & 78                                           
     & 91.23                                        & 364               
                                & 91.27                                 &
1365                                               & 91.26              
                         & 4368                                         
     & 91.26                                        & 12376             
                                & 91.26                                 
      & 31824                                              & 91.26      
                                 & 75582                                
             & 91.26                                        \\ \hline
\end{tabular}}
\end{table}

Table~{\ref{tabcnn}} shows the recognition accuracy of
implementing the CNN accelerator with the proposed memristor crossbar nodes. The CIFAR-10 dataset has ten classes, with each class having 6000 images. The CNN used for testing includes two convolution layers, with average pooling \footnote{Average pooling means calculating the average of each patch of feature maps obtained from the convolution layer, which goes as input to the subsequent layers.} and dropout ratio of 0.5, and two dense layers. The
first layer uses the ReLu activation function, while the last layer is the
softmax function. It can be seen that at the minimum, using two 8-level memristors or {seven} 3-level memristors gives accuracy that matches with ideal case results of memristor having infinite states, i.e., 91.26\%. For reliability of operations, it is more appropriate to use crossbar nodes with more memristors and conductive states. Even if some of the states disappear due to aging, the impact on overall recognition accuracy will be lower. 
\deleted{In contrast, if the memristor under consideration has large variability in the conductive states, it will be more robust to use such memristors in a
fewer number of states but having cells that have many such memristors.} {Any conductance level of the memristor can be imagined as having a mean conductance value with a standard variance. If the variance is large, the conductance values can overlap and introduce larger errors. For reducing this impact, it is recommended to allow larger separations between the conductance levels, using more memristors per node with fewer {conductance} levels. This \deleted{can lead to}\added{increases} robustness to conductance variations as can be observed in the example of Table \ref{tabcnnnoise}, for an 8-M node with $L=3$ or $L_C=45$. }

\begin{table}[!ht]
\centering
\caption{{Recognition accuracy of using a CNN accelerator with proposed memristor crossbar nodes. The tests are performed on CIFAR-10 dataset. Random variability of 10\% in the conductance values. }}
\label{tabcnnnoise}
\scalebox{.6}{
\begin{tabular}{|p{.5cm}|p{.5cm}|p{.5cm}|p{.5cm}|p{.5cm}|p{.5cm}|p{.5cm}|p{.5cm}|p{.5cm}|p{.5cm}|p{.5cm}|p{.5cm}|p{.5cm}|p{.5cm}|p{.5cm}|p{.5cm}|}
\hline
\multicolumn{2}{|l|}{\textbf{1-M node}}                    & \multicolumn{2}{l|}{\textbf{2-M node}}                     & \multicolumn{2}{l|}{\textbf{3-M node}}                     & \multicolumn{2}{l|}{\textbf{4-M node}}                     & \multicolumn{2}{l|}{\textbf{5-M node}}                     & \multicolumn{2}{l|}{\textbf{6-M node}}                     & \multicolumn{2}{l|}{\textbf{7-M node}}                     & \multicolumn{2}{l|}{\textbf{8-M node}}                     \\ \hline
\textbf{$L_C=L$} & \textbf{ACC (\%)} & \textbf{$L_C$} & \textbf{ACC (\%)} & \textbf{$L_C$} & \textbf{ACC (\%)} & \textbf{$L_C$} & \textbf{ACC (\%)} & \textbf{$L_C$} & \textbf{ACC (\%)} & \textbf{$L_C$} & \textbf{ACC (\%)} & \textbf{$L_C$} & \textbf{ACC (\%)} & \textbf{$L_C$} & \textbf{ACC (\%)} \\ \hline
2                           & 82.5                  & 3                           & 84.14                 & 4                           & 84.69                 & 5                           & 85.37                 & 6                           & 85.99                 & 7                           & 88.2                  & 8                           & 90.13                 & 9                           & 90.22                 \\ \hline
3                           & 83.94                 & 6                           & 86.62                 & 10                          & 90.63                 & 15                          & 90.83                 & 21                          & 90.92                 & 28                          & 90.97                 & 36                          & 90.96                 & 45                          & 91.02                 \\ \hline
4                           & 84.83                 & 10                          & 90.55                 & 20                          & 91                    & 35                          & 90.94                 & 56                          & 90.99                 & 84                          & 90.81                 & 120                         & 91.02                 & 165                         & 90.91                 \\ \hline
5                           & 85.29                 & 15                          & 90.87                 & 35                          & 91.01                 & 70                          & 91                    & 126                         & 90.96                 & 210                         & 90.93                 & 330                         & 91.03                 & 495                         & 90.99                 \\ \hline
6                           & 87.3                  & 21                          & 90.98                 & 56                          & 90.99                 & 126                         & 91.01                 & 252                         & 90.8                  & 462                         & 91                    & 792                         & 90.97                 & 1287                        & 91.06                 \\ \hline
7                           & 89.11                 & 28                          & 90.83                 & 84                          & 90.98                 & 210                         & 90.9                  & 462                         & 91.01                 & 924                         & 91.05                 & 1716                        & 90.95                 & 3003                        & 90.91                 \\ \hline
8                           & 90.16                 & 36                          & 91.01                 & 120                         & 90.93                 & 330                         & 91.02                 & 792                         & 90.92                 & 1716                        & 90.87                 & 3432                        & 90.98                 & 6435                        & 90.9                  \\ \hline
9                           & 90.13                 & 45                          & 90.96                 & 165                         & 90.92                 & 495                         & 90.99                 & 1287                        & 91.01                 & 3003                        & 90.95                 & 6435                        & 91.04                 & 12870                       & 90.8                  \\ \hline
10                          & 90.64                 & 55                          & 91                    & 220                         & 91                    & 715                         & 90.95                 & 2002                        & 90.98                 & 5005                        & 90.92                 & 11440                       & 91.01                 & 24310                       & 90.89                 \\ \hline
11                          & 90.81                 & 66                          & 91.02                 & 286                         & 90.83                 & 1001                        & 90.95                 & 3003                        & 91                    & 8008                        & 90.92                 & 19448                       & 91.05                 & 43758                       & 90.99                 \\ \hline
12                          & 90.7                  & 78                          & 90.94                 & 364                         & 90.95                 & 1365                        & 90.9                  & 4368                        & 90.89                 & 12376                       & 90.9                  & 31824                       & 90.98                 & 75582                       & 91                    \\ \hline
\end{tabular}}
\end{table}

Table {\ref{tabcnnnoise}} shows the recognition accuracy of \deleted{implementing} the CNN accelerator with \added{the }proposed memristor crossbar nodes \deleted{taking into account}\added{considering} a 10\% variability in the conductance levels. The variation of conductance is added such that it is random, and \deleted{the distribution of conductance} at a~ level follows a Gaussian distribution with maximum standard deviation set at 10\% of the mean value. The tests are repeated 30 times, and the average accuracy is reported. It can be observed from Table {\ref{tabcnnnoise}} that the conductance variability \deleted{results in a reduction in}\added{reduces} the overall accuracy. An increase in the number of memristors shows higher robustness. \deleted{This is possibly because the learning algorithm did not optimise the weights to its ideal optimal convergent values. However, these values is still lower than that obtained at higher number of conductance levels and memristors per node.} {Here, the CNN is trained and assumed to have obtained optimal weights. These weights are translated to the conductance values of the memristor superresolution nodes. However, the variability of the conductance values moves the weights further away from optimal values.} {In addition, the changes in conductance can also move the combination of weights back closer to optimal values increasing the accuracy. For example, when ($L_C$=2, $M$=1) and ($L_C$=3, $M$=2), the recognition  accuracy is slightly higher under variations by 0.07\% and 0.04\%.} \deleted{On the other hand, as the superresolution nodes have large number of conductance levels, these conductance shifts can also in some cases move the combination of weights back closer to optimal values, thereby showing an increase in accuracy. As an example, this special case is observed when the number of levels are low ($L_C$=2, $M$=1) ($L_C$=3, $M$=2), the recognition  accuracy is slightly higher under variations by 0.07\% and 0.04\%.}

{The over fitting in neural networks can be reduced by applying dropout based regularization during training stages. Further, as the neural network size grows, the energy consumption increases in both test and training stages. To address this, dropouts can be applied in both test and training stages, as shown in \cite{alexj2020}. The superresolution nodes  can use the same control circuits typically required to program a regular crossbar. In addition, by connecting the row-wise inputs together, superresolution nodes can be created without physically changing the regular crossbar design.}\deleted{The proposed crossbar nodes can be randomly switched off during the test and training stages to implement the dropout's, which can help reduce the issue of over-fitting and energy consumption \cite{alexj2020}. The crossbar nodes can also be increased in the conductance by feeding the same input to multiple rows. The practical realization of these super-resolution nodes is highly simplified with the controls performed at the input, without distributing the crossbar's physical configuration.}
This makes the approach universal to a wide range of analog neural network implementations, where the performance of large arrays gets affected by the variability of the conductance levels. 

The ability of the proposed approach to transforming any multi-level crossbar arrays to analog crossbar arrays offers improvement in precision and accuracy of the crossbar for analog computing. Any reduction in current errors implies higher accuracy of analog neural computations. Further, the superresolution crossbar nodes also provide increased fault tolerance and reliability. Even if any of the memristors fail or remain stuck at fault, the remaining memristors can still offer a high number of conductive states without compromising the analog neural network's overall performance.

\begin{figure}[!ht]
    \centering
    \includegraphics[width=80mm]{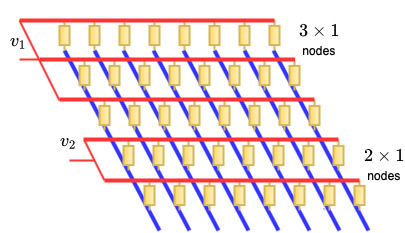}
    \caption{{Creating superresolution nodes with different values of $m$.}}
    \label{figmult}
\end{figure}

\paragraph*{Optimising superresolution crossbar size} {Figure 3 shows an example of a $2\times 1$ node within a crossbar. Figure \ref{figmult} shows an extension by combining a different number of rows to create superresolution nodes with different values of $m$. This approach can help to optimise the number of memristors required per crossbar for a given application. For example, if $3 \times 1$ nodes are used to create $2\times 8$ superresolution crossbar nodes, then a total of 48 memristors are required, while the combination of $3 \times 1$ and $2 \times 1$ nodes, as shown in Fig. \ref{figmult} requires a total of 40 memristors, resulting in lesser usage of crossbar area. To determine the value of $m$ for the superresolution node, for a fixed $L$ value, the mapping of $m$ with $L$ and $L_C$ can be applied. Suppose each memristor has $L=4$, and we want a conductance resolution of 0.1 (i.e., minimum 10 levels per node) and 0.01 (i.e., minimum 100 levels per node) in the crossbar. Then looking at Table III we can determine the value of $m$ for 0.1 as 2 (with $L_C=10$), and for 0.01 as 7 (with $L_C=120$). Such optimisation of crossbar will be useful for  applications where parts of the input signal vector within an input pattern such as attention or region of interest in images require higher precision of weights in neural network layers than others. Various applications that can be found for this approach is left as an open problem.}   

\section{Conclusion}

{\label{301895}}

In this paper, a compelling yet straightforward approach to improve the
performance and use of a memristive crossbar for analog neural network implementations are presented. The system shows how a conventional crossbar with memristor nodes with limited conductance resolution can be transformed into a high-resolution conductive node. The resolution mapping is obtained by input row mergers resulting in \(r\)-simplicial conductance sequences, with a non-linear growth in conductance with an increase in memristors per crossbar nodes. The increase in memristors per node, though it increases
the area, offers higher computation accuracy and a more straightforward mapping of weight values.

The reduced relative current errors with {an} increase in  memristors per crossbar nodes show improvement in precision and accuracy of the crossbar \deleted{multiply and accumulate}\added{MAC} calculations. This will, in particular, be useful in neural networks with {many} neurons layers, where even minor variations in the weights will have a more significant impact on inference accuracy. The current errors are not easy to compensate for during the inference stages and require additional fine-tuning. \deleted{Using the proposed super-resolution approach, the}\added{The} need for complex fine-tuning circuits during inference stages can be reduced \added{using the proposed super-resolution approach}. 

\section*{Acknowledgements}

 A. James's research is supported by industry research grant Ind/001/2020. L. Chua's  research is supported by  Grant FA9550-18-1-0016. 
\vspace{3in}
\end{document}